\documentclass[11pt]{article}
\usepackage{moriond,epsfig}

\def\al{\alpha}

\def\be{\begin{equation}}
\def\ee{\end{equation}}
\def\bea{\begin{eqnarray}}
\def\eea{\end{eqnarray}}

\def\P{{\rm I\kern-.15em P}}
\def\rel{$p_{T}^{\rm{rel}}$}
\def\q2{$Q^2$}
\begin{document}
\vspace*{4cm}
\title{BEAUTY PRODUCTION AT HERA}

\author{ K.H. KLIMEK }

\address{Institute of Experimental Physics, Hamburg University,\\
Luruper Chaussee 149, 22671 Hamburg, Germany}

\maketitle\abstracts{
The results of measurement of beauty quark production at HERA performed by the H1 and ZEUS
collaborations are presented. Total cross sections and the differential 
cross sections have been measured in both the photoproduction and the deep inelastic 
scattering regimes. The results are compared with the NLO QCD calculations. }

\section{Introduction}
%\subsection{Producing the Hard Copy}\label{subsec:prod}

Measurement of the production of heavy quarks can be used to probe Quantum Chromodynamics (QCD) 
and to understand the structure of the proton and photon.
The large beauty ($b$) quark mass
provides a hard scale that ensures perturbative QCD (pQCD) is applicable
and makes such calculations more reliable than for light quarks.
Additionally, the production of $b$ quarks is a source of background for many searches 
for new physics at existing and future experiments. However, for most measurements, 
the measured $b$-production cross section lies significantly above
QCD expectations calculated up to next-to-leading order (NLO)
\mbox{in the strong coupling constant, $\al_s$.}

Experimental procedures rely on signatures or features characteristic 
of $b$ production and decay. One of them is a measurement of the transverse 
momentum, \rel, of the muon produced in the semi-leptonic decay with respect to the
axis of the closest jet. The fact that this spectrum is harder for $b$ quarks then 
for $c$ or light quarks allows a statistical separation of the signal and the 
background. Also used in the signal extraction is a lifetime 
measurement provided by silicon vertex detectors. 
A further method used in this paper is tagging of both $D^*$ mesons and muons coming from 
the $b$ decays. This method is sensitive in the region of low $b$ quark transverse 
momentum and has a low background in the case where the $D^*$ and muon have opposite charge and 
lie in similar region of phase space.

The results obtained at HERA  are reviewed for the photoproduction 
($Q^2 \sim 0$, where $Q^2$ is the exchanged photon virtuality) and Deep 
Inelastic Scattering (DIS: $Q^2 > 1$ GeV$^2$) regions.

\section{Open Beauty in Photoproduction}

For photoproduction, where the exchanged photon has small \q2, the hadronic 
structure of the photon can be revealed (resolved photon). Therefore photoproduction 
at HERA is similar to a hadron collider and supplies complementary information. 
Using an integrated luminosity of 15 $\rm{pb^{-1}}$ the H1 collaboration 
used both \rel~and lifetime information to measure 
the total visible cross section in the reaction with a muon $\mu$ in the final 
state~\cite{H1:php:97}. For the kinematic range  
\q2 $< 1$ GeV$^2$, $0.1 < y < 0.8$, $p_T^\mu > 2 $ GeV, 
$35^\circ <  \theta^\mu < 130^\circ$ the result is  
$\sigma (ep \rightarrow b\bar{b} X \rightarrow \mu X) = 160 \pm 16\; \rm{(stat.)}\; \pm 
29 \; \rm{(syst.)}$ pb. The NLO QCD prediction, using the calculation implemented 
in the program FMNR~\cite{fmnr} with the Peterson fragmentation function~\cite{peterson} 
is $54 \pm 9$ pb, where the error corresponds to the uncertainties due to variation of the 
renormalisation and factorisation scale, and to the fragmentation. The NLO QCD prediction 
\mbox{is well below the data.}

\begin{figure}
%\rule{5cm}{0.2mm}\hfill\rule{5cm}{0.2mm}
%\vskip 2.5cm
%\rule{5cm}{0.2mm}\hfill\rule{5cm}{0.2mm}
\vspace*{-6.5cm}
\hspace*{-2.5cm}\psfig{figure=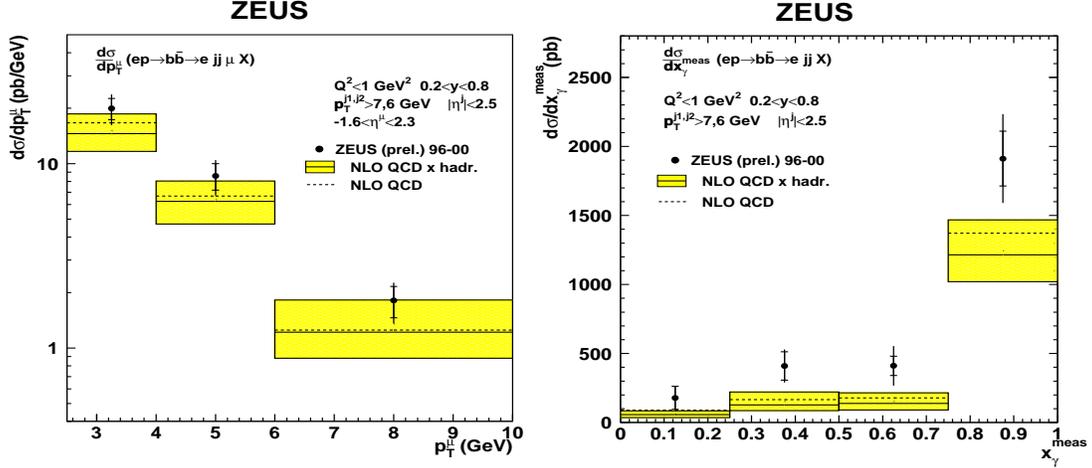,bbllx=0, bblly=0, bburx=567, bbury=650 , width=19cm, height=21.cm}
\vspace*{-9.cm}
\caption{Differential cross section of beauty photoproduction as 
a function of the muon transverse momentum (left)
and as a fraction of the photon momentum taking part in a hard process (right), 
compared with NLO QCD prediction.
\label{fig:ZEUSphp}}
%\vspace*{-1cm}
\end{figure}

The ZEUS experiment has measured differential cross sections of
beauty photoproduction using events with at least two jets  and a muon in
the final state \cite{ZEUS:php}. The fraction of events from $b$
decays has been extracted using the \rel~method.
The kinematic region is defined by \mbox{$Q^2 < 1$ GeV$^2$,}
$0.2<y<0.8$, $p_T^{\rm{Jet1(2)}} > 7(6)$ GeV, $|\eta^{\rm{Jet1(2)}}|<2.5$,
$p_T^\mu > 2.5$ GeV and \mbox{$-1.6<\eta^\mu<2.3$.}
\footnote{$\eta=-\ln(\tan\theta/2)$ is the pseudorapidity, where
$\theta$ is the polar angle measured with respect to the proton beam direction.}
\mbox{Figure \ref{fig:ZEUSphp}} shows a comparison between the measured
differential cross section as a function of a $p_T^\mu$ and a NLO QCD calculation. 
The QCD prediction was calculated using the FMNR program with
the hadronisation modeled by a Peterson function and 
the spectrum of the semi-leptonic muon momentum taken from JETSET~\cite{PYTHIA}.
The bands around the NLO prediction show the results obtained by
varying the 
$b$ quark mass and the renormalisation and factorisation scales.
The measured cross sections are a factor 1.4 larger than the central
NLO prediction but compatible with it within the experimental and
theoretical uncertainties.
In addition, a dijet cross section
$\sigma (ep \rightarrow b\overline{b}X \rightarrow~jet~jet~X)$
has been determined using PYTHIA~\cite{PYTHIA}
to extrapolate to the unmeasured part of the muon kinematics and to correct
for the branching ratio. For this measurement a different data sample has
been used with looser cuts on the transverse momentum of
the muon at large pseudorapidities. The result is
$\sigma^{\rm{dijet}} = 733 \pm 61 \pm 104$ pb while the NLO QCD prediction
is $381^{+117}_{-78}$ pb, which is a factor of two below the data.
In the same kinematic range the differential cross section as a function
of $x_\gamma^{\rm{meas}}$, 
which measures the fraction of the photon energy that takes part in the hard interaction, 
was calculated and is shown in Fig.~\ref{fig:ZEUSphp} together with the NLO QCD 
prediction described above. A clear contribution from the resolved photon
component ($x_\gamma \ll 1$) is seen and the agreement 
between the data and NLO QCD is comparable for the resolved and direct photon components.

\section{Open Beauty in DIS}

\begin{figure}
%\rule{5cm}{0.2mm}\hfill\rule{5cm}{0.2mm}
%\vskip 2.5cm
%\rule{5cm}{0.2mm}\hfill\rule{5cm}{0.2mm}
\vspace*{-5.cm}
\hspace*{-2.5cm}\psfig{figure=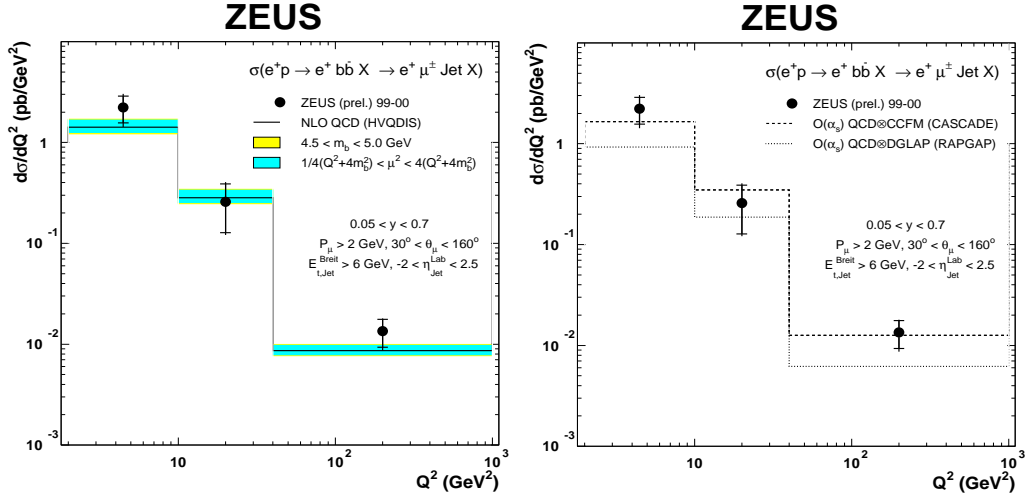,bbllx=0, bblly=0, bburx=567, bbury=650 , width=18cm, height=22.5cm}
\vspace*{-10.75cm}
\caption{Differential cross section of beauty production in DIS as a function $Q^2$ compared to NLO QCD calculations (left) and to CASCADE and RAPGAP Monte Carlo generators (right).
\label{fig:ZEUSdis}}
%\vspace*{-1cm}
\end{figure}

The ZEUS collaboration has measured differential cross sections as a function 
of $Q^2$ and $x$ for $b$ production in DIS, using the \rel~method. 
Events were selected by
requiring the presence of at least one muon in the final state
and at least one jet in the Breit frame~\footnote{In the Breit frame, defined by
$\vec{\gamma}+2x\vec{P} = \vec{0}$, where $\vec{\gamma}$ is the momentum of the
exchanged photon, $x$ is the Bjorken scaling variable and $\vec{P}$ is the
proton momentum, a purely space-like photon and a proton
collide head-on.}.
A total visible cross section of
$\sigma (ep \rightarrow e b\overline{b}X \rightarrow e~jet~\mu~X) = 
38.7\pm 7.7^{+6.1}_{-5.0}$ pb
was measured in the kinematic region defined by:
$Q^2 > 2$ GeV$^2$, $0.05<y<0.7$, $p^\mu > 2$ GeV, $30^\circ <\theta^\mu < 160^\circ$
and one jet in the Breit frame with $E_T^{\rm{Breit}}> 6$ GeV and $-2 < \eta^{Lab} < 2.5$.
This result has been compared with a NLO QCD calculation implemented
in the HVQDIS program~\cite{hvqdis}, after folding the
$b$ quark momentum spectrum with a Peterson fragmentation function 
and subsequently with a spectrum of the semi-leptonic muon momentum taken from JETSET~\cite{PYTHIA}.
The NLO QCD prediction is $28.1^{+5.3}_{-3.5}$ pb which agrees
with the measured value within the uncertainties. 
The prediction of the MC program CASCADE~\cite{cascade},
which implements a calculation based on the CCFM evolution equations~\cite{ccfm}
and uses a $k_T$-dependent gluon density, is 35 pb which
is in good agreement with the measurement.
In addition, the simulation gives a good description of the measured differential
cross sections.
The differential
cross section as a function of $Q^2$ compared to the NLO calculation and CASCADE and 
RAPGAP~\cite{rapgap} MC generators is shown in Fig. \ref{fig:ZEUSdis}.

The H1 collaboration measured the 
total visible cross section in the reaction with a muon in the final 
state~\cite{H1:php:97} using both \rel~and lifetime information. The kinematic 
range is defined by $2 <$ \q2 $< 100$ GeV$^2$, $0.05 < y < 0.7$ and $p_T^\mu > 2 $GeV, 
$30^\circ <  \theta^\mu < 160^\circ$, yielding  
$\sigma (ep \rightarrow b\bar{b} X \rightarrow \mu X) = 39 \pm 8\; \rm{(stat.)}\; \pm 
10 \; \rm{(syst.)}$ pb. The NLO QCD prediction from HVQDIS
is $11 \pm 2$ pb, where the error corresponds to the uncertainties due to variation of the 
renormalization or factorization scale, and to the fragmentation. The NLO QCD prediction 
is well below the data. The prediction from the CASCADE MC is 15 pb and 
also underestimates the measurement.

\section{$D^*-\mu$ Correlations}

Another method of separating charm and beauty contributions (the so-called ``double-tag
me -   \\thod'') 
exploits the angle and charge correlations of the $D^*$
meson and of the muon from the reaction
$ep\rightarrow eb\overline{b}X\rightarrow eD^*\mu X$.
The most interesting is the configuration
in which the muon and the $D^*$ originate from the same parent
$B$ meson and lead to unlike-charge sign $D^*-\mu$ pairs
produced in the same hemisphere.
 
Using the double-tag method and performing a likelihood fit on the
kinematic distributions, the H1 collaboration has extracted the visible cross sections
of beauty and charm production in the kinematic region defined by
$p_T^{D^*}>1.5$ GeV, $|\eta^{D^*}|<1.5$, $p_T^\mu > 1$ GeV,
$|\eta^\mu|<1.74$ and $0.05<y<0.75$\cite{H1:1016}.
The measured values are respectively \mbox{$\sigma^b=380\pm 120\pm 130$ pb} and
\mbox{$\sigma^c=720\pm 115\pm 245$ pb}.
The measurement is well above LO+parton shower MC expectations.
 
The ZEUS collaboration has performed a similar analysis optimising the selection
 for decays of $b$ quarks~\cite{ZEUS:784}.
The beauty cross section, measured in a slightly different
phase space ($p_T^{D^*}>1.9$ GeV, $|\eta^{D^*}|<1.5$, $p_T^\mu > 1.4$ GeV,
$-1.75<\eta^\mu<1.3$), is  $\sigma^b=214\pm 52^{+96}_{-84}$ pb.
The result agrees with the H1 measurement after applying the same cuts.
To compare the measured cross section with NLO QCD predictions of FMNR
 a photoproduction sample has been selected
by applying the cuts $Q^2 < 1$ GeV$^2$ and $0.05<y<0.85$. Additionally,
the measurement has been restricted to a $b$ quark rapidity range $\zeta^b <1$
where the $p_T$ and $\zeta^b$ distributions in the MC program used to extrapolate,
agree with the respective FMNR spectra to within $\pm15 \%$.
The result for the extrapolated cross section is
$\sigma(\gamma p \rightarrow b(\bar{b})X)=15.1\pm 3.9 ^{+3.8}_{-4.7}$ nb,
while the NLO prediction of FMNR is $5.0^{+1.7}_{-1.1}$ nb.

\section{Summary and Outlook} 

The understanding of the $b$ production mechanism still remains
a big puzzle in QCD.
A set of visible cross sections of beauty production have been
measured at HERA, both in the photoproduction and DIS regimes.
The comparison of the measurements and NLO QCD calculations varies
from agreement within the experimental and theoretical
uncertainties \mbox{to significant discrepancies.}

The HERA collider has started a new phase of operation at
higher luminosities. The new H1 and ZEUS vertex detectors and tracking triggers
will enhance the  $b$-tagging capabilities and allow 
more precise and differential measurements to be made within the next few years.

\end{document}